\documentclass[aip,apl,twocolumn]{revtex4}
\usepackage{graphicx}
\usepackage{amsmath}

\begin{document}

\title{A simple method to fabricate Josephson junctions}

\author{Imran Mahboob}
\email{imran.mahboob@ntt.com}
\affiliation{NTT Basic Research Laboratories, NTT Inc., 3-1 Morinosato-Wakamiya, Atsugi, Kanagawa 243-0198, ~Japan}

\author{Satoshi Sasaki}
\affiliation{NTT Basic Research Laboratories, NTT Inc., 3-1 Morinosato-Wakamiya, Atsugi, Kanagawa 243-0198, ~Japan}

\author{Takaaki Takenaka}
\affiliation{NTT Basic Research Laboratories, NTT Inc., 3-1 Morinosato-Wakamiya, Atsugi, Kanagawa 243-0198, ~Japan}

\begin{abstract}
{A minimal method to fabricate Al/AlO$_x$/Al Josephson junctions (JJs) using photolithography and argon etching, before metallization and oxidation, is demonstrated. JJs with areas ranging from 1 to 6 $\mu$m$^2$ can be fabricated and, with the appropriate oxidation conditions, the junction resistance can be varied by $\sim$2 orders of magnitude. Transmission electron microscopy reveals the successful fabrication of JJs with few grain boundaries suggesting reduced energy loss from two-level-systems. Superconducting QUantum Interference Devices (SQUIDs) fabricated from this methodology exhibit reduced resistance variation over multiple chips, compared with electron beam lithography, and the devices can sustain repeated thermal cycles to 10 mK with the excellent flux response remaining unchanged. The quantum applications of this technology are demonstrated by embedding a SQUID resonator into a 3D cavity and parametrically amplifying low photon numbers with gains of $\sim$40 dB. This work establishes the simplest approach to fabricating JJs to date, and it could prove pivotal to the widespread utilization of superconducting circuit-based quantum technologies.}
\end{abstract}

\maketitle

Josephson Junction (JJ) \cite{J1} based superconducting qubits \cite{J2} are at the forefront in the development of a quantum computer \cite{J3, J4} with the integration of large number of qubits in a single chip \cite{J5, J8}, quantum error correction beyond break-even \cite{J6} and millisecond qubit lifetimes \cite{J7}. Moreover, JJs play a pivotal role in many of the ancillary components needed in the realization of a practical quantum computer including quantum limited amplification \cite{J9} with Josephson parametric or travelling wave amplifiers \cite{jpa3, jpa8}, tunable filters \cite{J10}, circulators \cite{J13} and isolators \cite{J11, J12, J26}. 

Invariably JJs are fabricated via the Dolan bridge technique where shadow aluminum evaporation is interleaved with oxidation \cite{J16}. Although widely adopted, this method can result in stray JJs and it precludes aggressive cleaning in the proximity of the JJ, either via argon etching or oxygen plasma, as it can damage the resist bridge resulting in impaired junction fabrication. In recent times, the Manhattan technique has emerged \cite{J17} and it is especially suited to the fabrication of small area JJs with high resistance which are inherent to the most successful class of superconducting qubit: the transmon \cite{J27}.

\begin{figure}[!ht]
\vspace{0.1cm}
\includegraphics[scale=0.88]{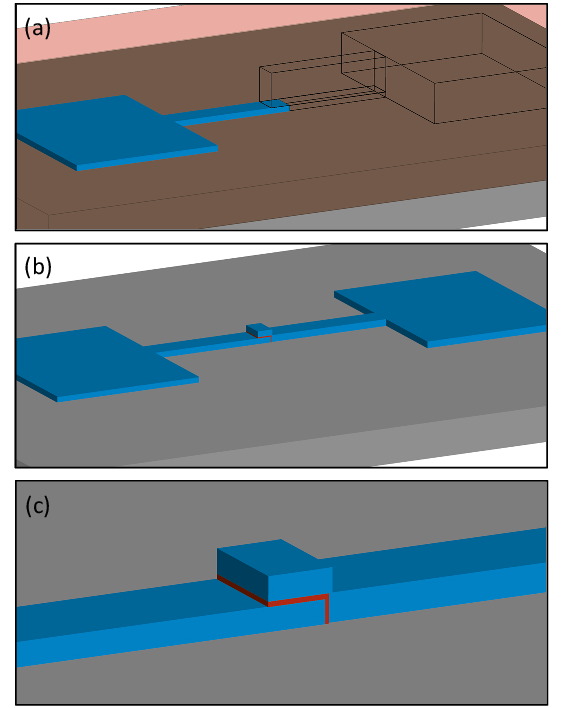}
\vspace{0.1cm}
\small{\caption{(a) The aluminum pattern of the base layer (blue) is deposited on a silicon substrate. Photoresist (pink) is patterned with 365nm UV light, and the resultant pattern is depicted by the hollow boxes (black lines). The overlap between the resist pattern and the base aluminum layer defines the JJ. (b) The deposition of the top metal layer (blue) is preceded by argon milling to expose pristine aluminum which is then oxidized (red) where the aluminum sandwich defines the JJ. A zoom of the junction region where the vertical component of the JJ makes an insignificant contribution to the overall junction resistance. }}
\end{figure}

\begin{figure*}[!ht]
\vspace{-0.1cm}
\includegraphics[scale=0.92, angle=-90]{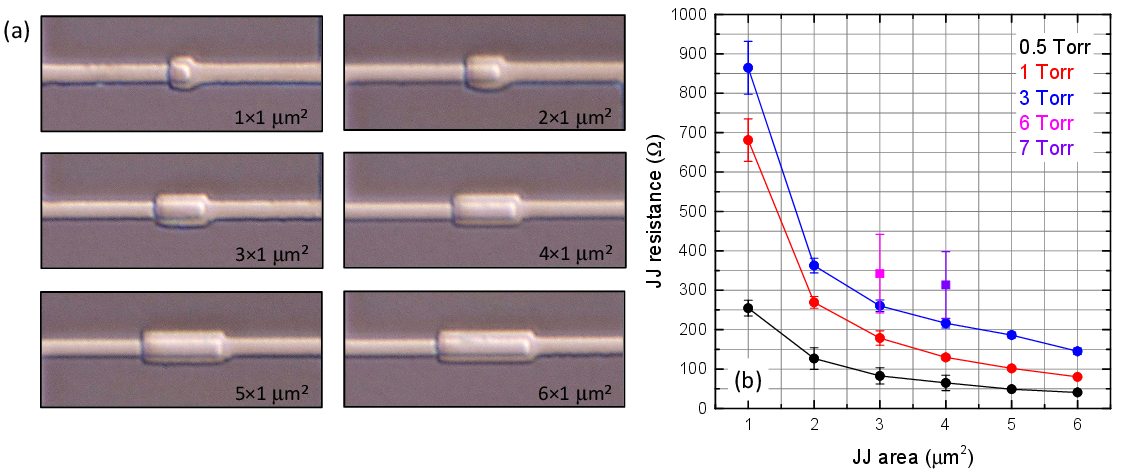}
\vspace{-0.1cm}
\small{\caption{(a) Optical micrographs, acquired with Keyence VHX-7000 digital microscope, reveal JJs from this simple fabrication method can be successfully fabricated, even with an area as small as $1 \times 1$ $\mu$m$^2$, with excellent alignment between base and top aluminum layers. (c) Room temperature JJ resistance as a function of junction size for test patterns (circles) and functional SQUID resonator devices (squares) all with 10 minutes oxidation with a range of pressures. All error bars correspond to 1 standard deviation and are invisible for the largest junctions.}}
\end{figure*}

Although a number of alternative JJ fabrication techniques have been developed \cite{J14, J15}, only the approach pioneered by Martinis et al. has received further attention \cite{J18}. Specifically, the JJ in this work consists of Al/AlO$_x$/Al sandwich deposited on a silicon or sapphire substrate where the oxidation is preceded in-situ with argon etching of the aluminum base layer. The subsequent device is then defined via reactive ion etching (RIE) of a photolithography or electron beam (e-beam) defined pattern in resist. This approach ensures that the JJ is clean without any atmospheric or resist related contamination and it can result in improved device performance. 

Subsequently this concept has been refined into the {\it overlap junction method} by Pappas et al. where base layer metallization is performed using the lift-off technique using a bi-layer resist stack that is patterned by e-beam \cite{J19}. Top layer metallization is also performed by using the lift-off technique where again a bi-layer resist stack is pattered with e-beam lithography. However, before aluminum deposition of the top layer, argon etching is performed to expose a pristine aluminum surface that is then controllably oxidized. Subsequently, this method was extended to photolithography which also employed a bi-layer resist stack and the lift-off technique, but it introduced a number of extra steps thus complicating the fabrication process \cite{J20}. However, due to improved device performance, the overlap technique has been extensively employed with photolithography to fabricate both qubits \cite{J21} and Josephson Parametric Amplifiers (JPAs) \cite{J22} where RIE is employed to define the pattern instead of lift-off. Intreastingly, this approach has also evolved into large-scale fabrication of JJs using 300mm silicon wafers \cite{J23} and recently qubits from this methodology have even been extended to a complementary metal-oxide semiconductor pilot line using industrial fabrication methods \cite{J24, J25}.

In contrast, here a minimal approach is developed that eliminates RIE and bi-layer resist stacks and instead employs a standard photolithography technique in combination with metal lift-off to fabricate JJs ranging from 1 to 6 $\mu$m$^2$ where junction resistance can be tuned from below 50 $\Omega$ to $\sim1$ k$\Omega$ with appropriate oxidation conditions. The utility of this approach is demonstrated with Superconducting QUantum Interference Devices (SQUIDs) that exhibit excellent flux response without any distortions over multiple thermal cycles and JPAs that can achieve gains of 40 dB with quantum limited noise.

The fabrication process starts with a $17 \times 18$ mm undoped silicon (100) chip with a resistivity of 20 k$\Omega$cm. The chip is first cleaned with organic solvents (acetone, isopropyl alcohol (IPA) and ethanol) followed by a diluted hydrofluoric (HF) acid dip (1 part $50\%$ HF:10 parts water) for 30 seconds. The chip is then washed in ultra-pure water and blown dry with nitrogen gas. Next iP3650 photoresist is spin coated with a thickness of 1 $\mu$m and is baked on a hotplate for 3 minutes at $100^{\circ}$ \cite{J28}. The resist is then patterned using Nano Systems Solutions DL-1000i maskless photolithography system with a 365 nm LED light source \cite{J29}. The patterns are then developed using TMA508 \cite{J30} for bare silicon or AZ developer \cite{J31} when the substrate has been metalized with aluminum. The developed chip is then rapidly placed into the load lock of a PLASSYS evaporator (MEB 550 S2-I UHV) and is vacuum pumped for several hours \cite{J32}. The chip is then transferred to the deposition chamber and is argon etched with a flow rate of 7 sccm for 60 seconds with an acceleration voltage of 120V and a beam current of 120 mA which removes $\sim15$ nm of both the silicon substrate, the photoresist and in the process all atmospheric contaminants. This is followed by deposition of 120 nm of $99.999\%$ pure aluminum via electron beam heating in ultra-high vacuum. Finally, the aluminum is capped via controlled oxidation in-situ, 90 Torr for 10 minutes, to protect from adsorption of atmospheric contaminants. The chip is removed from the PLASSYS system and is left overnight in a beaker of Microposit 1165 remover at room temperature \cite{J33}. Lift-off is executed the following day with acetone and IPA spray thus completing metallization of the base aluminum layer.

The metalized chip is then spin coated with iP3650 again and photolithography of the top layer is executed as schematically depicted in Fig. 1(a) where an overlap between the metalized base layer and the photo-pattern defines the JJ. The chip is again loaded into PLASSYS, as detailed earlier, and argon milling is carried out to not only remove atmospheric contamination on the silicon substrate but also from the exposed aluminum of the base layer where typically 20 nm of metal is removed. The Josephson tunnel junction is then synthesized by oxidizing the pristine aluminum and by controlling pressure and time, the desired junction resistance can be achieved. Finally, 120 nm of aluminum is deposited, which is again capped, and then lift-off is carried out as detailed earlier. A schematic of the fabricated JJ device is shown in Figs. 1(b) and 1(c).

Initially to calibrate the junction resistance and to understand the fidelity of this approach, test junctions were patterned with the area varied from 1-6 $\mu$m$^2$ in 1 $\mu$m$^2$ increments where the vertical junction size was kept fixed at 1 $\mu$m. Fifty test junctions of each size were fabricated in addition to a shorted pattern, all on the same chip, to enable the JJ resistance to be determined. The 0.5 $\mu$m resolution and layer alignment accuracy of Nano Systems Solutions DL-1000i maskless photolithography machine successfully fabricated overlap JJs as shown in the optical micrographs in Fig. 2(a). The resultant test junction's resistance was measured in a probe station, for a number of oxidation conditions, and the results are summarized in Fig. 2(b). All oxidation conditions used 10 minutes, and the mean junction resistance could be varied from below 50 $\Omega$ to almost 1 k$\Omega$ with error bars corresponding to 1 standard deviation, approximately $\pm 5-8\%$ of the mean resistance. Although this variation is somewhat larger than $\pm$2-5$\%$ variation achieved with e-beam lithography for similar patterns, there is enormous scope to improve this by using orthogonal alignment between base and top aluminum layers instead of the lateral alignment used here \cite{J21} 

\begin{figure}[!ht]
\vspace{0.0cm}
\includegraphics[scale=0.88, angle=-90]{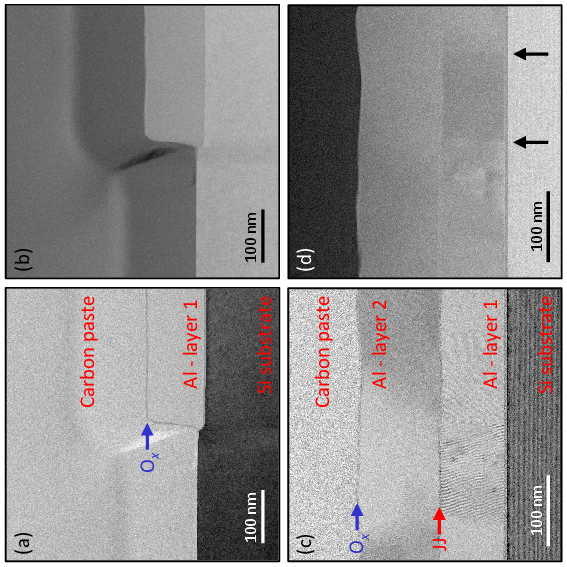}
\vspace{0.1cm}
\small{\caption{(a) and (b) BF and HAADF STEM with only the base 120 nm thick aluminum (Al) layer deposited on the argon etched silicon (Si) substrate. (c) and (d) BF and HAADF STEM following the completed device with the highly uniform 1-2 nm thick JJ clearly resolvable (red arrow). Also visible in the BF STEM images is the $\sim$2 nm thick oxide (O$_x$) cap (blue arrows). The carbon paste is employed as a conductive adhesive layer during focused ion beam milling used to access the JJ's cross-section for STEM. }}
\end{figure}

\begin{figure*}[!t]
\vspace{-0.1cm}
\includegraphics[scale=0.92, angle=-90]{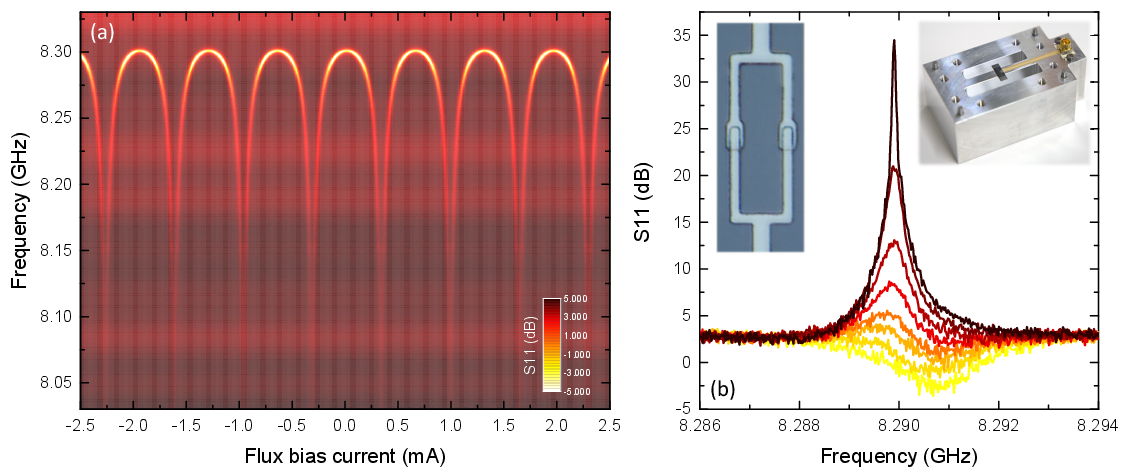}
\vspace{-0.1cm}
\small{\caption{(a) A SQUID based 2D resonator's response to magnetic flux bias is measured via a capacitively coupled 3D cavity at millikelvin temperatures. This device was repeatedly measured with more than 10 thermal cycles in a 6-month period, and it sustained a nominally unchanged flux response thus indicating the stability of the JJs in the SQUID. (b) Parametric amplification is measured with a flux bias of 0.16 mA that is combined with microwave modulation at $\sim 16.6$ GHz with increasing power from -33 to -26 dBm (orange to black lines) yielding a gain of almost 40 dB. The cavity is continuously probed with approximately 10 photons where the bare response (with microwave flux modulation deactivated) is shown in the yellow line. The left inset shows an optical micrograph of the SQUID fabricated from the simple method outlined in the main text. The SQUID is integrated into a 2D resonator which is then mounted inside and capacitively coupled to a pair of 3D cavities (only the lower half of the cavities is shown) as depicted in the right inset. The SQUID's flux response and parametric amplification from only the smaller 8.3 GHz cavity is shown. }}
\end{figure*}

The test patterns are instructive in guiding the oxidation conditions to achieve the desired junction resistance, however they are unrepresentative of functional devices which are patterned across the entire chip rather than at the center of the chip, for optimal resolution, as in the case of the test patterns. Specifically, SQUIDs with a pair of JJs embedded in a 2D resonator were fabricated using the methodology outlined above with 5 devices, in addition to a shorted reference device, patterned across the entire $17 \times 18$ mm chip with 4 such chips fabricated simultaneously. In this case, 2 junction sizes and 2 oxidation conditions were investigated and the resultant mean junction resistance with 1 standard deviation error bars are shown in Fig. 2(b). Unsurprisingly, the resistance variation for the functional device has increased to $\pm25\%$ of the mean resistance. However, to put this performance into context, it needs to be compared to the same devices fabricated by e-beam lithography which yield a mean resistance that is $150\%$ greater than the target value with 1 standard deviation error bars of almost $\pm200\%$ in 12 devices over 4 chips. This enormous distribution in the e-beam devices can be understood as a consequence of the Dolan bridge technique where small fluctuations in e-beam current and subsequent development are translated into variations in the resist bridge which are then massively amplified in the resultant JJ area spread from the shadow evaporation of aluminum. Consequently, this simplified fabrication protocol represents an enormous improvement compared to e-beam fabrication both in terms of ease in fabrication and the reduced variation in functional devices.

In order to evaluate the JJs fabricated from this simple method, Scanning Transmission Electron Microscopy (STEM) is carried out in both Bright-Field (BF) and High-Angle Annular Dark-Field (HAADF) modes. Figs. 3(a) and 3(b) show BF and HAADF STEM images respectively when only the first 120nm thick aluminum layer is deposited following argon etching. These images reveal that the silicon substrate is etched by approximately 15 nm, ensuring aluminum deposition occurs on a contaminant free substrate. Figs. 3(c) and 3(d) show BF and HAADF STEM images respectively after deposition of the top aluminum layer. From comparison of the thickness of the base and top aluminum layers, approximately 20 nm of aluminum is removed by the argon etch. In addition, the JJ of approximately 1-2 nm thickness can easily be identified and is further confirmed by the aluminum grain boundaries in the bottom aluminum layer not propagating to the top aluminum layer, see black arrows in Fig. 3(d). Finally, the STEM reveals a general lack of grain boundaries in both aluminum layers, which can act as source of dielectric energy loss in superconducting quantum circuits stemming from Two-Level-Systems (TLSs), and it indicates the high quality of the fabricated JJs \cite{J35}.

Lastly, the Al/AlO$_x$/Al JJs are measured at millikelvin temperatures by integrating them into a SQUID (as detailed above and shown in the left inset of Fig. 4(b)) which is embedded in a 2D resonator that is mounted inside a 3D cavity (see right inset in Fig. 4(b)) and is detailed in ref. \cite{J34}. The 2D resonator is capacitively coupled to the 3D cavity and it enables the SQUID's response to DC magnetic flux to be measured via the cavity's frequency modulation as shown in Fig. 4(a). This measurement reveals a highly symmetric and periodic SQUID response to magnetic flux indicating that the underlying JJs are identical. In addition, no flux jumps are observed or any coupling to parasitic TLSs, even with large area 3 $\mu$m$^2$ JJs which are normally susceptible to these parasitic loss channels \cite{J18}. Most pleasingly, 20 devices were measured which all yielded nominally identically flux response with one device measured over a period of 6 months, with more than 10 thermal cycles, exhibiting an almost unchanged response indicating the reproducibility and stability of the JJs fabricated from this simple method. 

Flux modulation at microwave frequencies can also enable parametric amplification \cite{J9, jpa4} and is investigated via the 3D cavity, as detailed in previous work, using the photolithographic SQUID resonator \cite{J34}. Specifically, the 8.3 GHz 3D cavity is excited with an average of 10 photons whilst simultaneously the magnetic flux is modulated at twice the cavity frequency. In this configuration, the cavity's frequency response is amplified with increasing microwave flux power, yielding a maximal gain of almost 40 dB as shown in Fig. 4(b). A comparison of the signal-to-noise ratio improvement and the underlying noise floor with previously measured e-beam devices enables half a photon of added noise to be inferred \cite{J34} thus demonstrating quantum limited performance being available to these devices.

Tantalizingly, this method has scope to be extended to the fabrication of qubits. Specifically employing orthogonal alignment between the base and top aluminum layers for 1 $\mu$m$^2$ JJs in combination with more aggressive oxidation conditions has the possibility of achieving the necessary resistance one would need for most species of superconducting qubits. Indeed, the work of Braum{\"u}ller et al. demonstrated that oxidizing 1 $\mu$m$^2$ JJs for 60 minutes at 23.6 Torr can yield the necessary resistance needed for transmon qubits \cite{J21}. As a result, this simple approach to fabricating JJs has the potential to make superconducting qubits, in addition to JPAs and JJ enabled ancillary technologies, becoming more accessible thus opening up the exciting possibility of quantum enabled technologies being ubiquitously employed.

A simple protocol to fabricate JJs is demonstrated using photolithography and the lift-off method. Both oxidation and metallization are preceded by argon etching to ensure contaminant free interfaces. The resulting junctions are structurally uniform without any defects. Patterning functional devices, over multiple chips, exhibits reduced JJ resistance variation in comparison to identical devices patterned by e-beam, via the Dolan bridge technique, thus improving both device yields and reproducibility. Indeed, measurements of SQUIDs fabricated from this protocol sustain an excellent flux response with no evidence of TLSs and this behavior remains unchanged, over multiple devices and thermal cycles, indicating the stability of the underlying JJs. Finally, a JPA implemented from the SQUID yields excellent gain with quantum limited noise thus confirming this fabrication methodology as being applicable to quantum technologies. Indeed, usage of slightly more aggressive oxidation conditions will bring this protocol in-line for qubit fabrication, and it has the possibility of making superconducting circuit-based quantum technologies becoming more ubiquitously available.

The authors thank Aijiro Saito for supporting the device fabrication, Dr Kosuke Kakuyanagi for maintaining the PLASSYS system and Dr Yui Ogawa for assisting with the optical microscopy.

\bibliography{Fab}

@article{J1,
author = {A. L. Solovjev and S. I. Bondarenko},
collaboration = {},
title = {On the sixtieth anniversary of the discovery of the non-stationary Josephson effect},
year = {2024},
journal = {Low Temp. Phys.},
volume = {50},
pages = {921-924},
}

@article{J2,
author = {P. Krantz and  and M. Kjaergaard and F. Yan and T. P. Orlando and S. Gustavsson and W. D. Oliver},
collaboration = {},
title = {A quantum engineer's guide to superconducting qubits},
year = {2019},
journal = {Appl. Phys. Rev.},
volume = {6},
pages = {021318},
}

@article{J3,
author = {Frank Arute and others},
collaboration = {},
title = {Quantum supremacy using a programmable superconducting processor},
year = {2019},
journal = {Nature},
volume = {574},
pages = {505-510},
}

@article{J4,
author = {He-Liang Huang and Dachao Wu and Daojin Fan and Xiaobo Zhu},
collaboration = {},
title = {Superconducting quantum computing: a review},
year = {2020},
journal = {Sci. China Inf. Sci.},
volume = {63},
pages = {180501},
}

@article{J5,
author = {M. AbuGhanem},
collaboration = {},
title = {IBM quantum computers: evolution, performance, and future directions},
year = {2025},
journal = {J. Supercomput.},
volume = {81},
pages = {687},
}

@article{J6,
author = {V. V. Sivak and A. Eickbusch and B. Royer and S. Singh and I. Tsioutsios and S. Ganjam and A. Miano and B. L. Brock and A. Z. Ding and L. Frunzio and S. M. Girvin and R. J. Schoelkopf and M. H. Devoret},
collaboration = {},
title = {Real-time quantum error correction beyond break-even},
year = {2023},
journal = {Nature},
volume = {616},
pages = {50-55},
}

@article{J7,
author = {Matthew P. Bland and others},
collaboration = {},
title = {Millisecond lifetimes and coherence times in 2D transmon qubits},
year = {2025},
journal = {Nature},
volume = {647},
pages = {343-348},
}

@article{J8,
author = {Harald Putterman and others},
collaboration = {},
title = {Hardware-efficient quantum error correction via concatenated bosonic qubits},
year = {2025},
journal = {Nature},
volume = {638},
pages = {927-934},
}

@article{J9,
author = {J. Aumentado},
collaboration = {},
title = {Superconducting parametric amplifiers: The state of the art in Josephson parametric amplifiers},
year = {2020},
journal = {IEEE Microwave magazine},
volume = {21},
pages = {45-59},
}

@article{J10,
author = {S. Kono and K. Koshino and D. Lachance-Quirion and A. F. van Loo and Y. Tabuchi and A. Noguchi and Y. Nakamura},
collaboration = {},
title = {Breaking the trade-off between fast control and long lifetime of a superconducting qubit},
year = {2020},
journal = {Nature Commun.},
volume = {11},
pages = {3683},
}

@article{J11,
author = {Baleegh Abdo and Nicholas T. Bronn and Oblesh Jinka and Salvatore Olivadese and Antonio D. Corcoles and Vivekananda P. Adiga and Markus Brink and Russell E. Lake and Xian Wu and David P. Pappas and Jerry M. Chow},
collaboration = {},
title = {Active protection of a superconducting qubit with an interferometric Josephson isolator},
year = {2019},
journal = {Nature Commun.},
volume = {10},
pages = {3154},
}

@article{J12,
author = {F. Lecocq and L. Ranzani and G. A. Peterson and K. Cicak and X. Y. Jin and R. W. Simmonds and J. D. Teufel and J. Aumentado},
collaboration = {},
title = {Efficient Qubit Measurement with a Nonreciprocal Microwave Amplifier},
year = {2019},
journal = {Phys. Rev. Lett.},
volume = {126},
pages = {020502},
}

@article{J13,
author = {Benjamin J. Chapman and Eric I. Rosenthal and Joseph Kerckhoff and Bradley A. Moores and Leila R. Vale and J. A. B. Mates and Gene C. Hilton and Kevin Lalumiere and Alexandre Blais and K. W. Lehnert},
collaboration = {},
title = {Widely Tunable On-Chip Microwave Circulator for Superconducting Quantum Circuits},
year = {2019},
journal = {Phys. Rev. X},
volume = {7},
pages = {041043},
}

@article{J14,
author = {Florent Lecocq and Ioan M Pop and Zhihui Peng and Iulian Matei and Thierry Crozes and Thierry Fournier and Cecile Naud and Wiebke Guichard and Olivier Buisson},
collaboration = {},
title = {Junction fabrication by shadow evaporation without a suspended bridge},
year = {2011},
journal = {Nanotechnology},
volume = {2},
pages = {315302},
}

@article{J15,
author = {J. T. Monroe and D. Kowsari and K. Zheng and C. Gaikwad and J. Brewster and D. S. Wisbey and K. W. Murch},
collaboration = {},
title = {Optical direct write of Dolan–Niemeyer-bridge junctions for transmon qubits},
year = {2021},
journal = {Appl. Phys. Lett.},
volume = {119},
pages = {062601},
}

@article{J16,
author = {G. J. Dolan},
collaboration = {},
title = {Offset masks for lift‐off photoprocessing},
year = {1977},
journal = {Appl. Phys. Lett.},
volume = {31},
pages = {337-339},
}

@article{J17,
author = {A. Potts and P. R. Routley and G. J. Parker and J. J. Baumberg and P. A. J. de Groot},
collaboration = {},
title = {Novel fabrication methods for submicrometer Josephson junction qubits},
year = {2001},
journal = {J. Mater. Sci.: Mater. Electron.},
volume = {12},
pages = {289-293},
}

@article{J18,
author = {Matthias Steffen and M. Ansmann and R. McDermott and N. Katz and Radoslaw C. Bialczak and Erik Lucero and Matthew Neeley and E. M. Weig and A. N. Cleland and John M. Martinis},
collaboration = {},
title = {State Tomography of Capacitively Shunted Phase Qubits with High Fidelity},
year = {2006},
journal = {Phys. Rev. Lett.},
volume = {97},
pages = {050502},
}

@article{J19,
author = {X. Wu and J. L. Long and H. S. Ku and R. E. Lake and M. Bal and D. P. Pappas},
collaboration = {},
title = {Overlap junctions for high coherence superconducting qubits},
year = {2017},
journal = {Appl. Phys. Lett.},
volume = {111},
pages = {032602},
}

@article{J20,
author = {Mustafa Bal and others},
collaboration = {},
title = {Overlap junctions for superconducting quantum electronics and amplifiers},
year = {2021},
journal = {Appl. Phys. Lett.},
volume = {118},
pages = {112601},
}

@article{J21,
author = {Jochen Braumuller and Joel Cramer and Steffen Schlor and Hannes Rotzinger and Lucas Radtke and Alexander Lukashenko and Ping Yang and Sebastian T. Skacel and Sebastian Probst and Michael Marthaler and Lingzhen Guo and Alexey V. Ustinov and Martin Weides},
collaboration = {},
title = {Multiphoton dressing of an anharmonic superconducting many-level quantum circuit},
year = {2021},
journal = {Phys. Rev. B},
volume = {91},
pages = {054523},
}

@article{J22,
author = {Patrick Winkel and others},
collaboration = {},
title = {Nondegenerate Parametric Amplifiers Based on Dispersion-Engineered Josephson-Junction Arrays},
year = {2020},
journal = {Phys. Rev. Applied},
volume = {13},
pages = {024015},
}

@article{J23,
author = {N. Foroozani and others},
collaboration = {},
title = {Development of transmon qubits solely from optical lithography on 300 mm wafers},
year = {2019},
journal = {Quantum Sci. Technol.},
volume = {4},
pages = {025012},
}

@article{J24,
author = {J. Van Damme and S. Massar and R. Acharya and Ts. Ivanova and D. Perez Lozano and Y. Canvel and M. Demarets and D. Vangoidsenhoven and Y. Hermans and J. G. Lai and A. M. Vadiraj and M. Mongillo and D. Wan and J. De Boeck and A. Potocnik and K. De Greve},
collaboration = {},
title = {Advanced CMOS manufacturing of superconducting qubits on 300 mm wafers},
year = {2024},
journal = {Nature},
volume = {634},
pages = {74-79},
}

@article{J25,
author = {S. J. K. Lang and T. Mayer and J. Weber and C. Dhieb and Eisele and W. Lerch and Z. Luo and C. Moran Guizan and E. Music and L. Sturm-Rogon and D. Zahn and R.N. Pereira and C. Kutter},
collaboration = {},
title = {CMOS-compatible processing and room-temperature characterization at the wafer level for scalable quantum computing},
year = {2025},
journal = {Phys. Rev. Applied},
volume = {24},
pages = {054052},
}

@article{J26,
author = {C. S. Kow and M. T. Bell},
collaboration = {},
title = {Traveling-Wave Parametric Amplifier with Passive Reverse Isolation},
year = {2026},
journal = {Phys. Rev. X},
volume = {16},
pages = {021003},
}

@article{J27,
author = {Jens Koch and Terri M. Yu and Jay Gambetta and A. A. Houck and D. I. Schuster and J. Majer and Alexandre Blais and M. H. Devoret and S. M. Girvin and R. J. Schoelkopf},
collaboration = {},
title = {Charge-insensitive qubit design derived from the Cooper pair box},
year = {2007},
journal = {Phys. Rev. A},
volume = {76},
pages = {042319},
}

@misc{J28,
howpublished = {\url{https://www.tok-pr.com/en/products/photoresist/g-i-line.html}},
}

@misc{J29,
howpublished = {\url{https://www.nanosystem-solutions.com/en/product/maskless}},
}

@misc{J30,
howpublished = {\url{https://www.kanto.co.jp/products/denshi/kinosei/resist_strippers_developer/tmk_tma.html}},
}

@misc{J31,
howpublished = {\url{https://www.microchemicals.com/AZ-Developer-5.00-l/1000001}},
}

@misc{J32,
howpublished = {\url{https://plassys.com/categories/MEB-Ebeam/MEB550S}},
}

@misc{J33,
howpublished = {\url{https://www.agasem.com/products-and-services/semiconductors/microposit-remover-1165/}},
}

@article{J34,
author = {Imran Mahboob and others},
collaboration = {},
title = {A three-dimensional Josephson parametric amplifier},
year = {2022},
journal = {Appl. Phys. Express},
volume = {15},
pages = {062005},
}

@article{J35,
author = {Janka Biznarova and Amr Osman and Emil Rehnman and Lert Chayanun and Christian Krizan, Per Malmberg and Marcus Rommel and Christopher Warren and Per Delsing and August Yurgens and Jonas Bylander and Anita Fadavi Roudsari},
collaboration = {},
title = {Mitigation of interfacial dielectric loss in aluminum-on-silicon superconducting qubits},
year = {2024},
journal = {Npj Quantum Inf.},
volume = {10},
pages = {78},
}

@article{jpa3,
author = {M. A. Castellanos-Beltran and K. W. Lehnert},
collaboration = {},
title = {Widely tunable parametric amplifier based on a superconducting quantum interference device array resonator},
year = {2007},
journal = {Appl. Phys. Lett.},
volume = {91},
pages = {083509},
}

@article{jpa4,
author = {T. Yamamoto and K. Inomata and M. Watanabe and K. Matsuba and T. Miyazaki and W. D. Oliver and Y. Nakamura and J. S. Tsai},
collaboration = {},
title = {Flux-driven Josephson parametric amplifier},
year = {2008},
journal = {Appl. Phys. Lett.},
volume = {93},
pages = {042510},
}

@article{jpa8,
author = {C. Macklin and K. O'Brien and D. Hover and M. E. Schwartz and V. Bolkhovsky and X. Zhang and W. D. Oliver and I. Siddiqi},
collaboration = {},
title = {A near-quantum-limited Josephson traveling-wave parametric amplifier},
year = {2015},
journal = {Science},
volume = {350},
pages = {307-310},
}

\end{document}